\documentclass[a4paper, 11pt]{article}
\usepackage[english]{babel}
\makeatletter

\usepackage{geometry,graphicx,color}
\usepackage{amsfonts, amsmath, amssymb, slashed,dsfont}
\usepackage{tensor}
\usepackage[pdftex]{hyperref}
\usepackage{epsfig}
\usepackage{pifont}
\usepackage{enumitem}
\usepackage{bm}
\usepackage{mathrsfs} 
\usepackage[hyperref,thmmarks,amsmath]{ntheorem}
\usepackage{amsmath,amssymb,slashed}
\usepackage{tikz}
\usetikzlibrary{positioning}
\usetikzlibrary{decorations.markings}
\usetikzlibrary{arrows.meta}

\numberwithin{equation}{section}
\newcommand\bbone{\mathbb{I}}

\newcommand{\Jb}{\overline{J}}

\newcommand\phib{\overline{\phi}}

\newcommand\sstar{\text{\ding{86}}}

\newcommand{\tops}[2]{\texorpdfstring{#1}{#2}}

\newcommand{\institute}[1]{\newcommand{\@institute}{#1}}
\renewcommand{\maketitle}{
\vspace*{0.5\baselineskip}
{
\center\LARGE\noindent\@title\par
}%
\vspace{1.5\baselineskip}
{
\center\normalsize\noindent\ignorespaces\@author\par
}%
\vspace{0.5\baselineskip}
{
\center\normalsize\ignorespaces\@institute\par
}%
\vspace{2\baselineskip}
}%

\begin{document}

\title{Field theories on $\rho$-deformed Minkowski space-time}
\author{Kilian Hersent, Jean-Christophe Wallet}
\institute{%
\textit{IJCLab, Universit\'e Paris-Saclay, CNRS/IN2P3, 91405 Orsay, France}\\%
\bigskip
e-mail:  
\href{mailto:kilian.hersent@universite-paris-saclay.fr}{\texttt{kilian.hersent@universite-paris-saclay.fr}}

\href{mailto:jean-christophe.wallet@universite-paris-saclay.fr}{\texttt{jean-christophe.wallet@universite-paris-saclay.fr}}
}%
\maketitle

\begin{abstract} 
We study one-loop perturbative properties of scalar field theories on the $\rho$-Minkowski space. The corresponding star-product, together with the involution are characterized from a combination of Weyl quantization and defining properties of the convolution algebra of the Euclidean group linked to the coordinate algebra of the $\rho$-Minkowski space. The natural integration measure linked to the Haar measure of the Euclidean group defines a trace for the star-product. One-loop properties of the 2-point and 4-point functions for families of complex-valued scalar field theories on $\rho$-Minkowski space are examined. For scalar theories with orientable interaction, the 2-point function is found to receive UV quadratically diverging one-loop corrections in 4 dimensions while no IR singularities generating UV/IR mixing appears. These however occur in the one-loop corrections to the 4-point function. As well, one-loop 2-point functions for theories with non-orientable interaction involve such IR singularities. These results are discussed.

\end{abstract}

\newpage
\section{Introduction}
\label{sec:intro}

\paragraph{}
There is a rather widely accepted consensus that Quantum Gravity \cite{zerevue} should give rise to a quantum space-time at some effective regime. Various quantum space-times, conveniently modeled in the framework of noncommutative geometry \cite{connes}, have been considered in the physics literature for a long time. Among them, those on which acts a deformed Poincar\'e symmetry usually encoded in a Hopf algebra, which is interpreted as the quantum analog of space-time symmetry, are regarded as physically promising. The deformation parameter they involve is often assumed to be a new universal constant, possibly related to the Planck mass.\\

Quantum space-times with "Lie algebra noncommutativity" have a preeminent place in the physics literature devoted to the different approaches to Quantum Gravity and related (noncommutative) field theories \cite{szabo} and gauge theories \cite{physrep}. Among them is the very popular $\kappa$-Minkowski space introduced more than 30 years ago \cite{luk-ruegg}, \cite{majid1}. It has been the subject of a huge literature \cite{luk-rev} in view of its possible physical interest, providing in particular a realisation of the Double Special Relativity \cite{amel-nature}, \cite{kow-glik} or in relationship with Relative Locality \cite{rel-loc1, rel-loc2, rel-loc3}. Recall that the $\kappa$-Minkowski space is linked by a duality to a deformation of the Poincar\'e algebra called the $\kappa$-Poincar\'e (Hopf) algebra. Field theories as well as gauge theories on this quantum space have been studied \cite{marija11}-\cite{MW2020a}. $\mathbb{R}^3_\lambda$ is another interesting quantum space based on an $\mathfrak{su}(2)$ noncommutativity. Fields theories, which are known to have in particular relationships with a class of brane models \cite{schomerus} as well as with group field theory models \cite{gft} have also received interest, \cite{R1}-\cite{R7}.\\

Another deformation of the Minkowski space, called the $\rho$-Minkowski space, which also is acted on by a deformation of the Poincar\'e algebra, called the $\rho$-Poincar\'e algebra, has received some attention more recently, albeit having been considered in an interesting algebraic study a long ago \cite{luk-wor}. See also \cite{infinimany}. Its physical relevance lies in its possible relation to Relative Locality \cite{rel-loc1} and possible emergence in black-hole physics \cite{marija1}, \cite{marija2}. The concepts of localizability and quantum observers have been examined in \cite{localiz1} (see also \cite{localiz2}). Informally, the $\rho$-Minkowski space can be viewed as generated by the following Lie algebra of coordinates
\begin{align}
    [x_0, x_1] = i \rho x_2, &&
    [x_0, x_2] = - i \rho x_1, &&
    [x_1, x_2] = 0,
    \label{coord-alg-intro}
\end{align}
where $\rho$ has the dimension of a length which is supplemented by another generator $x_3$ which is central{\footnote{Notice that one could interchange $x_0$ and $x_3$ which would not alter the result for the star-product, apart from a mere change of notations but would correspond to a physically different situation where the time $x_0$ would stay "commutative". In the following, we will not consider this possibility, thus staying with a "noncommutative time".}}. The non-trivial part of \eqref{coord-alg-intro} is the Euclidean algebra $\mathfrak{e}(2)$.\\
Some quantum properties of a real-valued (massive) scalar field theory on the $\rho$-Minkowski space with quartic interaction has been examined in \cite{rho3}{\footnote{See also \cite{rho4}.}}, focusing on the 2-point function at the one-loop order. In this work, the star-product modeling the deformation of the Minkowski space is obtained from a Drinfeld twist. One salient conclusion of this work is that in 4 dimensions, UV/IR mixing occurs.\\

The purpose of the present paper is to extend the above work by examining the one-loop properties of the 2- and 4-point functions of complex-valued scalar field theories on $\rho$-Minkowski space with orientable or non-orientable quartic interactions. Recall that an orientable interaction is such that the field and its conjugate, says $\phi$ and $\phi^\dag$, alternate (and the converse for a non-orientable interaction). The star-product we will use is obtained by adapting the construction carried out in the case of the $\kappa$-Minkowski space \cite{DS}, \cite{PW2018} and applied to \cite{PW2019}, \cite{MW2020a}. It is thus different from the one on which are based \cite{marija1}, \cite{rho3}. It is obtained by adapting the construction used in the case of the $\kappa$-Minkowski space \cite{DS, PW2018} inherited from the old works \cite{vonNeum, Weyl}. We stress that this is a natural construction of a star-product on $\rho$-Minkowski in view of the common structures of the groups underlying respectively $\kappa$- and $\rho$-Minkowski quantum spaces. The mathematical equipment is relatively modest and is presented in Section \ref{sec:section2}. The resulting star-product is used in Section \ref{se:section3} to study one-loop properties of complex-valued scalar field theories with orientable or non-orientable quartic interactions. In Section \ref{sec:section4}, we summarize and discuss the results and list interesting issues to be examined.\\

\section{Star-product for \tops{$\rho$}{rho}-Minkowski from Weyl quantization}
\label{sec:section2}
\paragraph{}
In this section, we will present the construction of a star-product for the $\rho$-Minkowski space. The method we will follow is in fact very natural and is actually inherited from pioneering works of von Neumann and Weyl \cite{vonNeum, Weyl}.
In modern language, it combines properties of the convolution product defining the group algebra linked to the noncommutative coordinates algebra with the Weyl quantization operator\footnote{
This applied to the Heisenberg algebra yields the Moyal product, interpretable as a twisted convolution product.}.
This framework has been applied to obtain a convenient star-product for the celebrated $\kappa$-Minkowski space \cite{DS, PW2018}. This latter was further exploited to explore quantum properties of noncommutative scalar field theories and of gauge theories on this quantum space \cite{PW2019, MW2020a}. For more mathematical details, see e.g.\ in \cite{physrep}.

\subsection{General set-up}
\label{subsec:21}
\paragraph{}
It is instructive to give a general construction which exploit standard properties of harmonic analysis of semi-direct products of (locally compact) groups, a mathematically interesting type of groups to which pertain the affine group and the Euclidean group, respectively related to the $\kappa$-Minkowski and $\rho$-Minkowski spaces. Both groups have the following semidirect product structure
\begin{equation}
    \mathcal{G}
    := H\ltimes_{{\phi}} \mathbb{R}^{n}
    \label{genesemidirect}
\end{equation}
$n\ge1$, where $H$ is a subgroup of $GL(n,\mathbb{R})$ and the (continous) morphism $\phi: H \to \mathrm{Aut}(\mathbb{R}^{d})$ is defined by the usual action of any matrix in $H\subset GL(n,\mathbb{R})$ on elements of $\mathbb{R}^n$. This is simply given by
\begin{equation}
    \phi_a(x) = a x,
    \label{phiaction}
\end{equation}
for any $a\in H$, $x\in\mathbb{R}^n$. In \eqref{genesemidirect}, $\mathbb{R}^n$ is the additive group of real numbers. Recall that it is acted on by $\phi$ \eqref{phiaction} which will alter the structure of the group law, compared to an usual direct product. The actual relation with $\kappa$-Minkowski or $\rho$-Minkowski depends on the choice for $H$ which will be given in a while.

\paragraph{}

Denoting by $(a,x)$ the elements of $\mathcal{G}$, the structure of the group $\mathcal{G}$ is defined by
\begin{align}
    (a_1, x_1) (a_2, x_2)
    &= (a_1 a_2, x_1 + a_1 x_2), &
    \label{w1}\\
    (a,x)^{-1}
    &= (a^{-1}, -a^{-1} x), &
    \bbone_{\mathcal{G}}
    = (\bbone_H,0)
    \label{w2},
\end{align}
where the action of $\phi$ on the second group factor is explicit in the RHS of \eqref{w1}. To illustrate these relations and for further convenience, it may be useful to introduce the following faithful representation of $\mathcal{G}$, $\gamma:\mathcal{G}\to \mathbb{M}_{n+1}(\mathbb{C})$
\begin{equation}
    \gamma:(a,x)
    \longmapsto \begin{pmatrix} a & x \\ 0 & 1   \end{pmatrix}
\end{equation}
for any $a\in H$, $x\in\mathbb{R}^n$ (the matrix is blockwise). Note that the action of any element $\gamma((a,x))$ restricted on $\mathbb{R}^n$ is given by $\gamma((a,x))y=ay+x$ for any $y\in\mathbb{R}^n$.

\paragraph{}
Given a locally compact group $\mathcal{G}$, the related convolution product is defined by
\begin{equation}
    (F\circ G)(s)
    = \int_{\mathcal{G}}\ d\mu(t) F(s t) G(t^{-1})
    \label{convol-gene}
\end{equation}
for any $F,G\in L^1(\mathcal{G})$, $s\in\mathcal{G}$, where $d\mu(t)$ denotes the left-invariant Haar measure. Recall that this latter is related to the right-invariant Haar measure, says $d\nu$, by the expression $d\nu(s)=\Delta(s^{-1})d\mu(s)$ for any $s\in\mathcal{G}$, where the group homomorphism $\Delta:\mathcal{G}\to\mathbb{R}^+$ is called the modular function. Whenever the group is unimodular, one has $\Delta(s)=1$
for any $s\in\mathcal{G}$, so that the left-invariant and right-invariant measures coincide.

From general properties on Haar measures on semidirect products, the left-invariant Haar measure and modular function of a localy compact group of the form $\mathcal{G} := H\ltimes_{{\phi}} \mathbb{R}^{n}$ as given by \eqref{genesemidirect}, are respectively given by (in obvious notations)
\begin{equation}
    d\mu_\mathcal{G}((a,x))
    = d\mu_{\mathbb{R}^n}(x)\ d\mu_H(a)\ |\det(a)|^{-1}
    \label{measure},
\end{equation} 
and
\begin{equation}
    \Delta_\mathcal{G}((a,x))
    = \Delta_{\mathbb{R}^n}(x)\ \Delta_H(a)\ |\det(a)|^{-1}
    \label{modul-function}
\end{equation}
for any $a\in H$, $x\in\mathbb{R}^n$. Plainly, $d\mu_{\mathbb{R}^n}(x)$ is the Lebesgue measure on $\mathbb{R}^n$, usually noted $d^nx$, while $d\mu_H$ and $\det(a)$ depend on the choice of $H$ assumed here to be a subgroup of $GL(n,\mathbb{R})$ which will be fixed in a while.

\paragraph{}
The related algebra, which will play a central role in the ensuing construction, is known as the convolution algebra, denoted hereafter by $\mathbb{C}(\mathcal{G}):=(L^1(\mathcal{G}),\circ,^\sstar)$ which is a $\star$-algebra thanks to the natural involution defined by
\begin{equation}
    F^\sstar(x)
    = {\overline{F}}(x^{-1}) \Delta_\mathcal{G}(x^{-1})
    \label{involution}
\end{equation}
for any $F\in L^1(\mathcal{G})$, $x\in\mathcal{G}$, where ${\overline{F}}$ is the complex conjugate of $F$. Given a unitary representation of $\mathcal{G}$, says $\pi_U:\mathcal{G}\to\mathcal{B}({\mathcal{H}})$, the induced $\star$-representation of $\mathbb{C}(\mathcal{G})$ on $\mathcal{B}({\mathcal{H}})$, $\pi:\mathbb{C}(\mathcal{G})\to\mathcal{B}({\mathcal{H}})$, is defined by 
\begin{equation}
   \pi(F)
   = \int_\mathcal{G} d\mu_\mathcal{G}(x) F(x) \pi_U(x)
   \label{inducedrep} 
\end{equation}
for any $F\in\mathbb{C}(\mathcal{G})${\footnote{$F$ must have compact support. In addition, notice that $\pi_U$ must be strongly continuous, which will be the case in the following analysis.}} and is bounded and non-degenerate. Recall that
\begin{align}
    \pi(F\circ G)
    = \pi(F)\pi(G), &&
    \pi(F)^\ddag
    = \pi(F^\sstar)
    \label{pi-morph}
\end{align}
with $F=\mathcal{F}f$ and $G=\mathcal{F}g$ and $\pi(F)^\ddag$ denotes the adjoint operator of $\pi(F)$.

\paragraph{}
Now, assume that {\it{the elements of $\mathbb{C}(\mathcal{G})$ are functions on a momentum space}}, i.e. any $F\in\mathbb{C}(\mathcal{G})$ can be written as $F=\mathcal{F}f$ where $\mathcal{F}$ denotes the Fourier transform\footnote{
    Our convention for the Fourier transform is $\mathcal{F}f(p) = \int \frac{d^dx}{(2\pi)^d}\ e^{- i p x} f(x)$ and $f(x) = \int d^dp\ e^{i p x} \mathcal{F}f(p)$.
}.
This, combined with the Weyl quantization operator given by
\begin{equation}
    Q(f)
    = \pi(\mathcal{F}f)
    \label{weyl-operat}
\end{equation}
generates a product on the algebra of functions which are the inverse Fourier transform of the elements of $\mathbb{C}(\mathcal{G})$. In this respect, it is natural to interpret this algebra as an algebra of functions of space(-time) coordinates and the product mentioned just above as the star-product on the noncommutative (quantum) space modeled by this algebra. From \eqref{pi-morph} and \eqref{weyl-operat}, one easily obtains
\begin{align}
    Q(f\star g)
    = Q(f) Q(g), &&
    (Q(f))^\ddag
    = Q(f^\dag),
    \label{defining-relations},
\end{align}
from which follow
\begin{align}
    f\star g
    = \mathcal{F}^{-1} (\mathcal{F}f \circ \mathcal{F}g), && 
    f^\dag
    = \mathcal{F}^{-1} (\mathcal{F}(f)^\sstar).
    \label{star-prodetinvol}
\end{align}
For more details on this construction, see in \cite{physrep}.

\paragraph{}
The actual nature of the above quantum space depends on the particular choice of $H$ as a subgroup of $GL(n,\mathbb{R})$. This latter has many subgroups, each one possibly giving rise to a particular quantum space whose interest in physics should be examined.

\paragraph{}
It is instructive to close this subsection by briefly sketching the case $H=a\bbone_n$, $a>0$, which is isomorphic to $\mathbb{R}^+$, assuming in this case $n=d-1$, . This diagonal subgroup of $GL(n,\mathbb{R})$ corresponds to the affine group $\mathbb{R}^+\ltimes_\phi\mathbb{R}^{d-1}$ of $\mathbb{R}^{d-1}$ which is known to lead to the $d$-dimensional $\kappa$-Minkowski space, see e.g.\ \cite{DS, PW2018}. Indeed, all the results relative to the star-product and involution used in these latter references can be easily obtained by first setting $a=e^{-p_0/\kappa}$ in the above formulas, thus interpreting $p_0$ as the time-like component of the momentum. This combined with \eqref{phiaction}-\eqref{modul-function} gives immediately the left Haar measure and the modular funtion respectively given by $d\mu = d^{d-1}p\ dp_0\ e^{(d-1)p_0/\kappa}$ and $\Delta=e^{(d-1)p_0/\kappa}$ (signaling that the affine group is not unimodular), while the right Haar reduces to the usual Lebesgue measure. Focusing on the right Haar measure as in \cite{DS, PW2018}, one easily deduces the expression for the convolution product and involution, namely 
\begin{align}
    (\mathcal{F}f \circ \mathcal{F}g)(p_0, \vec{p})
    &= \int_{\mathbb{ R}^d} dq_0\ d^{d-1}q\ \mathcal{F}f \left(p_0 - q_0, p-e^{(q_0-p_0)/\kappa}\vec{q}\right) \mathcal{F}g(q_0, \vec{q})
    \label{kappa-convol}, \\
    \mathcal{F}f^\sstar(p_0, \vec{p})
    &= e^{(d-1)p_0/\kappa}\ \overline{\mathcal{F}f}(-p_0, -e^{p_0/\kappa}\vec{p})
    \label{kappa-invol}.
\end{align}
where $\overline{F}$ denotes the complex conjugation of $F$. Finally, the simple application of an inverse Fourier transform on \eqref{kappa-convol} and \eqref{kappa-invol} yields the expression for the star-product for the $\kappa$-Minkowski space derived in \cite{DS, PW2018}, together with the corresponding natural involution.

\subsection{Star product for \tops{$\rho$}{rho}-Minkowski space}
\paragraph{}
In the following, we will mainly consider a situation where $H\subset O(2)$, a case which is actually connected to the 4-dimensional $\rho$-Minkowski space considered in \cite{rho3}. We set $n = 2$. Indeed, the non-trivial part of the coordinate algebra is the Euclidean algebra $\mathfrak{e}(2)$, with Euclidean group $\mathbb{E}(2)=O(2)\ltimes_\phi\mathbb{R}^2$. Recall that the Euclidean group $\mathbb{E}(n)=O(n)\ltimes_\phi\mathbb{R}^n$ is the group of isometries of the $n$-dimensional euclidean space involving translations, rotations and reflections, where the (additive) group $\mathbb{R}^n$ is isomorphic to the translation group $\mathcal{T}(n)$.

\paragraph{}
From now on, we will focus on the orientation preserving isometries, thus assuming $H=SO(2)$ leading to the special Euclidean group 
\begin{equation}
    \mathcal{G}_\rho
    := SE(2)
    = SO(2) \ltimes_\phi \mathbb{R}^2
    \label{g-rho}. 
\end{equation}
Then, assuming as in Subsection \ref{subsec:21} that the group elements describe a momentum space, one can write any element of $\mathcal{G}_\rho$ as $( R(\rho p_0), \vec{p})$ with $\vec{p} \in \mathbb{R}^2$. Here, $R(\rho p_0)$ denotes a $2 \times 2$ rotation matrix with defining (dimensionless) parameter $\rho p_0$, where $\rho$ has inverse mass dimension $-1$, to be identified with the deformation parameter of the Minkowski space, and $p_0$ is identified with the time-like component of a momentum $(p_0, \vec{p})$. Notice that the present convention for the parameter $\rho$ is the same as the convention used e.g.\ in \cite{rho3} where the deformation parameter has also the dimension of a length, which physically may be identified with the Planck length.

\paragraph{}
Then, eqns.\ \eqref{w1} and \eqref{w2} take the form
\begin{align}
  (R(\rho{p_0}),\vec{p})\ (R(\rho q_0),\vec{q})
  &= \Big(R(\rho(p_0 + q_0)), \vec{p} + R(\rho{p_0})\vec{q} \Big), &
  \label{rho1} \\
  (R(\rho{p_0}), \vec{p})^{-1}
  &= \Big(R({-\rho p_0}), - R(-\rho{p_0})\vec{p} \Big), &
  \bbone = \bbone_2
  \label{rho2},
\end{align}
which characterize the structure of $\mathcal{G}_\rho$. Note that first order approximation of these group relations agrees with equation (3) of \cite{rel-loc3}.

\paragraph{}
It is known that $\mathcal{G}_\rho$ is unimodular which can be easily recovered from \eqref{modul-function}, owing to the unimodularity of $\mathbb{R}^2$ and $SO(2)$ and the fact that $\Delta_{H}(a)=|\det a|$ for any $a\in O(n)$. From \eqref{measure}, one easily realizes that the measure on $\mathcal{G}_\rho$ reduces to the usual Lebesgue measure. We therefore set as usual 
\begin{equation}
    d\mu 
    = d^2p\ dp_0
    = d^3p,
    \label{3dmesure}
\end{equation}
in obvious notations. The convolution product and involution can now be written as 
\begin{align}
    (\mathcal{F}f \circ \mathcal{F}g)(p_0, \vec{p})
    &= \int d^3q\ \mathcal{F}f \Big(R({\rho}({p_0+q_0})), \vec{p} + R({\rho}{p_0})\vec{q} \Big) \mathcal{F}g \Big(R({-{\rho}q_0}), - R({-{\rho}q_0}) \vec{q} \Big)
    \label{rho3} \\
    \mathcal{F}f^*(p_0,\vec{p})
    &= \overline{\mathcal{F}f} \Big(R({-{\rho}p_0}),-R({-{\rho}p_0})\vec{p} \Big),
    \label{rho4}
\end{align}
for any $\mathcal{F}f, \mathcal{F}g \in L^1(\mathcal{G}_\rho)$.

\paragraph{}
We are done. Indeed, by simply combining the various Fourier transforms in \eqref{rho3}, \eqref{rho4} with \eqref{star-prodetinvol}, one obtains the expressions for the star-product and related involution which define the noncommutative $\rho$-Minkowski space, namely 
\begin{align}
    (f\star_\rho g)(x_0,\vec{x}) 
    &= \int \frac{dp_0}{2\pi}\ dy_0\ e^{-i p_0 y_0} f(x_0 + y_0, \vec{x}) g(x_0, R(-\rho p_0)\vec{x}),
    \label{star-rho} \\
    f^\dag(x_0,\vec{x})
    &= \int \frac{dp_0}{2\pi}\ dy_0\ e^{-i p_0 y_0} \overline{f}(x_0 + y_0, R(-\rho p_0) \vec{x} ),
    \label{invol-rho}
\end{align}
for any $f, g \in L^1(\mathbb{R}^3)$. The above resulting associative $*$-algebra can be extended to a suitable multiplier algebra of tempered distributions as in the $\kappa$-Minkowski case \cite{DS}. Let $\mathcal{M}^3_\rho$ denotes this $^\star$-algebra.

\paragraph{}
Some comments are now in order.

First, it can be easily verified that \eqref{star-rho} leads to the following coordinate algebra
\begin{align}
    [x_0, x_1] = i \rho x_2, &&
    [x_0, x_2] = - i \rho x_1, &&
    [x_1, x_2] = 0,
    \label{coord-alg}
\end{align}
which is the non-trivial part of the coordinate algebra for the $\rho$-Minkowski space \cite{marija1, rho3}. The full algebra for $\rho$-Minkowski is obtained from \eqref{coord-alg} by supplementing the generators $x_0,x_1,x_2$ with a central element $x_3$. The extension of the star-product to incorporate this extra coordinate is straightforward.

\paragraph{}
Next, it can be easily verified that \eqref{star-rho} and \eqref{invol-rho} reduce respectively to the usual commutative product between functions and complex conjugation at the commutative limit $\rho\to 0$.

\paragraph{}
Then, one observes that the star-product \eqref{star-rho} is different from star-product used in \cite{rho3}. Instead, the structure of the integrand is rather close to the one of the star-product for the $\kappa$-Minkowski space derived in \cite{DS, PW2018}. In particular, observe the second argument of the rightmost function in \eqref{star-rho} which represents the spatial coordinates acted on by $SO(2)\simeq\mathbb{S}^1$. In the $\kappa$-Minkowski case, the $SO(2)$ action is replaced by the action of $\mathbb{R}^+$, as recalled in Subsection \ref{subsec:21}. This can be expected from the common overall structure of the groups underlying these star-products. Both groups are of the form \eqref{genesemidirect}, the only change stemming from the choice of the subgroup $H$ with corresponding change in the way the $\mathbb{R}^n$ group factor in \eqref{genesemidirect} is acted on.

\paragraph{}
The star-product for the $\rho$-Minkowski space derived above actually corresponds to a 3-dimensional situation, which is again apparent from the underlying group $SO(2)\ltimes_\Phi\mathbb{R}^2$, leading to one dimension for $SO(2)$ (as it is a one-parameter group) supplementing the obvious two dimensions for the second group factor. In order to cope with 4-dimensional situation, one way is to add a central element, says $x_3$, to the coordinate algebra, as done in \cite{marija1, rho3}. This is what we will do in the subsequent analysis. The corresponding extension of the star-product together with the corresponding involution are simply given by
\begin{align}
    (f\star_\rho g)(x_0,\vec{x},x_3)
    &=\int \frac{dp_0}{2\pi}\ dy_0\ e^{-i p_0 y_0} f(x_0 + y_0, \vec{x}, x_3) g(x_0, R(- \rho p_0) \vec{x}, x_3),
    \label{star-final} \\
    f^\dag(x_0, \vec{x}, x_3)
    &= \int \frac{dp_0}{2\pi}\ dy_0\ e^{- i p_0 y_0} \overline{f}(x_0 + y_0, R( - \rho p_0) \vec{x}, x_3),
    \label{invol-final}
\end{align}
for any $f,g\in\mathcal{M}^4_\rho$, the 4-dimensional extension of $\mathcal{M}^3_\rho$.

\paragraph{}
The natural measure to be used for $\mathcal{M}^3_\rho$ is the 3-d Lebesgue measure \eqref{3dmesure} as discussed at the beginning of this subsection which trivially extends to the 4-d measure in the case of the 4-dimensional $\rho$-Minkowski space $\mathcal{M}^4_\rho$. From now on, we will denote generically the multiplier algebra by $\mathcal{M}_\rho$, irrespective of the dimension of the space.

\paragraph{}
We end up this subsection by giving some properties of the star-product $\star_\rho$ which will be used in the next section.

First, it can be easily verified that the following formulas hold ($x=(x_0,\vec{x},x_3)$)
\begin{align}
    \int d^4x\ (f \star_\rho g^\dag)(x)
    = \int d^4x\ f(x) \overline{g}(x), &&
    \int d^4x\ f^\dag(x)
    = \int d^4\ \overline{f}(x),
    \label{formule1}
\end{align}
which imply that 
\begin{equation}
    \int d^4x\ (f\star_\rho f^\dag)(x)
    = \int d^4x\ f(x)\overline{f}(x)\ge0\label{norml2}
\end{equation}
for any $f,g\in\mathcal{M}_\rho$. One concludes from \eqref{norml2} that the integral $\int d^4x$ defines a positive map $\int d^4x:\mathcal{M}_{\rho}^{+}\to\mathbb{R}^+$ where $\mathcal{M}_{\rho}^+$ denotes the set of positive elements of $\mathcal{M}_{\rho}$.

\paragraph{}
It turns out that the Lebesgue integral $\int d^4x $ defines a trace w.r.t.\ the star-products \eqref{star-rho} and \eqref{star-final}. This trace is {\it{not}} twisted contrary to the natural trace arising in the description of the $\kappa$-Minkowski space recalled at the end of Subsection \ref{subsec:21}. Hence, the usual cyclicity holds, namely
\begin{equation}
 \int d^4x\ (f\star_\rho g)(x)=\int d^4x\ (g\star_\rho f)(x),\label{cyclic}
\end{equation}
for any $f,g\in\mathcal{M}_\rho$, which can be easily verified from an elementary calculation.

\paragraph{}
In order to build action functionals, we will need a Hilbert product. It is defined by
\begin{equation}
    \langle f,g\rangle :=\int d^4x\ (f^\dag\star_\rho g)(x)=\int d^4x\ \overline{f}(x)g(x),\label{zehilbertprod}
\end{equation}
for any $f,g\in\mathcal{M}_\rho$ where the rightmost equality stems from the combination of \eqref{formule1} and \eqref{cyclic}, which formally coincides with the usual $L^2$ product.

Besides, one can check that 
\begin{equation}
    (f\star_\rho g)^\dag(x)=(g^\dag\star_\rho f^\dag)(x)
\end{equation}
for any $f,g\in\mathcal{M}_\rho$.

\paragraph{}
At this stage, one comment is in order. One observes that the star-product $\star_\rho$ is {\it{stricto sensu}} not closed w.r.t.\ the trace $\int d^4x$, since one has $\int d^4x f\star_\rho g\ne\int d^4x\ fg$ for arbitrary complex-valued functions. However, this star-product becomes closed when the relevant set of functions is restricted to real-valued functions. Roughly speaking, $\star_\rho$ is not far from the closedness w.r.t.\ the trace.

It must be stressed that the above formulas are obviously valid in the 3-dimensional case.

\paragraph{}
As a final remark, the coordinate algebra \eqref{coord-alg} in ``Cartesian coordinates'' can be written in ``cylindrical coordinates'' via the change of variable $x_r = \sqrt{x_1^2 + x_2^2}$ and $x_\varphi = \exp\left(i \arctan(\frac{x_2}{x_1}) \right)$, $x_0$ and $x_3$ being unchanged. With these new coordinates, the relations \eqref{coord-alg} becomes
\begin{align}
    [x_0, x_\varphi] = \rho x_\varphi
    \label{coord-alg-cyl}
\end{align}
the other bracket being zero. One can then perform the same analysis as before with $H = U(1)$, the complex rotations. One obtains that rotations on $\mathbb{R}^2$ are now rotations of $U(1)$, explicitly $R(-\rho p_0)\vec{x}$ now corresponds to $e^{i\rho p_0} x_\varphi$.

The structure equation of $\mathcal{G}_\rho$ \eqref{rho1} and \eqref{rho2} then becomes
\begin{align}
    (e^{i \rho p_0}, p_\varphi)\ (e^{i \rho q_0}, q_\varphi)
    &= \big( e^{i \rho (p_0 + q_0)}, p_\varphi + e^{i \rho p_0} q_\varphi \big), && 
    \label{rho1-cyl} \\
    (e^{i \rho p_0}, p_\varphi)^{-1}
    &= (e^{ - i \rho p_0}, - e^{-i \rho p_0} p_\varphi), &
    \mathbb{I} &= 1.
    \label{rho2-cyl}
\end{align}
Finally, we obtain the star-product and involution
\begin{align}
    (f \star_\rho g)(x_0, x_r, x_\varphi, x_3)
    &= \int \frac{dp_0}{2\pi}\ dy_0\ e^{-i p_0 y_0} f(x_0 + y_0, x_r, x_\varphi, x_3) g(x_0, x_r, e^{i \rho p_0} x_\varphi, x_3),
    \label{star-final-cyl} \\
    f^\dag(x_0, x_r, x_\varphi, x_3)
    &= \int \frac{dp_0}{2\pi}\ dy_0\ e^{- i p_0 y_0} \overline{f}(x_0 + y_0, x_r, e^{i \rho p_0} x_\varphi, x_3).
    \label{invol-final-cyl}
\end{align}

In view of the correspondence $R(-\rho p_0)\vec{x} \to e^{i\rho p_0} x_\varphi$, the results of the section \ref{se:section3} will be the same regardless the coordinate choice. Therefore, we will stick to the Cartesian coordinates.

\paragraph{}
One should note that the commutation relation \eqref{coord-alg-cyl} is similar to the $1 + 1$-dimensional $\kappa$-Minkowski one performing the change $\rho \to \frac{i}{\kappa}$. The major difference being that here $x_\varphi$ is an angle and so has compact support. This can be traced back to the fact that $1 + 1$-d $\kappa$-Minkowski has matrix group $H = \mathbb{R}^+$, and cylindrical $\rho$-Minkowski has $H = U(1)$. Therefore, the second matrix group corresponds to a compactification of the first one. 

This similarity between $\rho$ and $\kappa$ can go further as, upon the change $\rho \to \frac{i}{\kappa}$ and the compactification, the star-product and involution \eqref{star-final-cyl} and \eqref{invol-final-cyl} exactly corresponds to the ones of $\kappa$-Minkowski obtained from \eqref{kappa-convol} and \eqref{kappa-invol}.

\section{Scalar field theories on \tops{$\rho$}{rho}-Minkowski space}
\label{se:section3}
\paragraph{}
In this section, we will perform a first exploration of one-loop properties of scalar field theories on $\rho$-Minkowski, paying attention to the possible occurrence of IR singularities in the 2-point functions which may signal UV/IR mixing. A more detailed analysis of the perturbative behaviour of the scalar theories will be published elsewhere.

\paragraph{}
We will consider mainly the following (positive) action in 4 dimensions
\begin{align}
\begin{aligned}
    S(\phi, \overline{\phi})
    &= \langle \partial \phi, \partial \phi \rangle
    + m^2 \langle \phi, \phi \rangle
    + g \langle \phi^\dag \star_\rho \phi, \phi^\dag \star_\rho \phi \rangle \\
    &= \int d^4x\ (\partial_\mu \overline{\phi} \partial_\mu \phi + m^2 \phib \phi)
    + g \int d^4x\ \phi^\dag \star_\rho \phi \star_\rho \phi^\dag \star_\rho \phi,
\end{aligned}
    \label{action1}
\end{align}
where the fields $\phi$, $\overline{\phi}$ and the parameter $m$ have mass dimension 1 and $g$ is a dimensionless coupling constant, thus restricting the interaction term to a so-called orientable interaction in the terminology of noncommutative field theories \cite{PW2019}. To obtain the expression in the RHS of \eqref{action1}, \eqref{zehilbertprod} has been used. Note that the formal commutative limit of the action \eqref{action1} coincides formally with an ordinary massive $\phi^4$ theory. From time to time, we will compare the results to those obtained from a non-orientable interaction term \eqref{nonorient-1}.

\paragraph{}
The perturbative expansion is obtained from the generating functional of the connected Green functions $W(J, \overline{J})$, namely one has
\begin{equation}
    e^{W(J, \overline{J})} 
    = \int D\phi\ D\phib\ e^{ - \big(S(\phi,\overline{\phi}) + S_s(J,\overline{J}) \big)},
\end{equation}
where the source term $S_s$ takes the form
\begin{equation}
    S_s(J,\overline{J})
    = \langle J, \phi \rangle + \langle \phi, J \rangle
    = \int d^4x\ \overline{J}\phi + J\overline{\phi}.
    \label{source}
\end{equation}
From the functional relation $W(J, \Jb) = W_0(J, \Jb) + \ln\big(1 + e^{- W_0(J, \Jb)} [e^{- S_{\mathrm{int}}} - 1] e^{W_0(J, \Jb)}\big)$, one infers that the relevant one-loop contributions are generated by
\begin{equation}
    W_{(1)}(J, \Jb)
    = W_0(J, \Jb) - e^{- W_0(J, \Jb)}\ S_{\mathrm{int}}\left(\frac{\delta}{\delta J}, \frac{\delta}{\delta\Jb}\right)\ e^{W_0(J,\Jb)}
    \label{generateur}
\end{equation}
up to an unessential additive constant, where $W_0(J,\Jb)$ is the free generating functional of the connected Green functions and $S_{\mathrm{int}}(\frac{\delta}{\delta J },\frac{\delta}{\delta\Jb})$ is obtained as usual from the interaction term in \eqref{action1} through the replacement $\phi\to\frac{\delta}{\delta\Jb}$, $\phib\to\frac{\delta}{\delta J}$.

\paragraph{}
The free generating functional $W_0(J,\Jb)$ is given by
\begin{equation}
    e^{W_0(J, \Jb)}
    = e^{\int d^4p\ \Jb(p) (p^2+m^2)^{-1} J(p)},
    \label{Wzero}
\end{equation}
where $J(p) = \int \frac{d^4x}{(2\pi)^4}\ e^{-ipx} J(x)$.

\paragraph{}
The quartic interaction term $S_ {\mathrm{int}}$ in the action \eqref{action1} can be cast into the form
\begin{equation}
    S_{\mathrm{int}}
    = (2\pi)^4 \int \Big(\prod_{j=1}^4 dk_j \Big)\ \phib(k_1) \phi(k_2) \phib(k_3) \phi(k_4)\ V(k_1, k_2, k_3, k_4),
    \label{csint}
\end{equation}
with again $\phi(p) = \int \frac{d^4x}{(2\pi)^4}\ e^{-ipx} \phi(x)$, where the vertex function is given by
\begin{align}
\begin{aligned}
   V(k_1, k_2, k_3, k_4)
   =&\; g\ \delta(k^0_1 - k^0_2 + k^0_3 - k^0_4) \delta(k^3_1 - k^3_2 + k^3_3 - k^3_4) \\
   &\times \delta^2 \Big( R(\rho k^0_1) (\vec{k}_1 - \vec{k}_2) + R(\rho k^0_4)(\vec{k}_3 - \vec{k}_4) \Big).
\end{aligned}
    \label{zebigvertex}
\end{align}
The two first delta's express the conservation laws for the energy and third component of the momentum which take the usual form. The last delta signals that the conservation law for the 1 and 2 components of the momentum are altered by the deformation. One observes that this "deformed" law bears some similarity with the corresponding law obtained for a similar field theory on $\kappa$-Minkowski in e.g.\ \cite{PW2018} (see formula (3.41) of this reference) with however the so-called modular factors $\sim e^{-p_0/\kappa}$ replaced by rotation operators. This can be expected in view of the semidirect product structure of each of the groups underlying the two noncommutative spaces, as discussed in Subsection \ref{subsec:21}.

One should note that this vertex has the symmetries
\begin{align}
    V(1234) = V(4321), &&
    V(1234) = V(2143).
    \label{eq:vertex_symm}
\end{align}

\begin{figure}
    \begin{minipage}{.249\textwidth}
         \centering
    \begin{tikzpicture}[scale = 1.2]
        \draw[black] (-.1,-.1) rectangle (.1,.1);
        \draw[black] (-.7, -.7) node[anchor= east]{$\phi (k_2)$} to (-.1, -.1);
        \draw[black] (-.7,  .7) node[anchor= east]{$\phib(k_1)$} to (-.1,  .1);
        \draw[-{To}, black] (.1, .1) to (.6, .6) to[out= 45, in= 90] (.9,0)
            node[anchor = east]{$k_3$};
        \draw[black] (.1,-.1) to (.6,-.6) to[out=-45, in=-90] (.9,0);
        \draw (0,-.8) node[anchor = north]{$V(1233)$} to (0,-.8);
    \end{tikzpicture}
    \end{minipage}%
    \begin{minipage}{.249\textwidth}
         \centering
    \begin{tikzpicture}[scale = 1.2]
        \draw[black] (-.1,-.1) rectangle (.1,.1);
        \draw[black] (.7,  .7) node[anchor= west]{$\phib(k_1)$} to (.1,  .1);
        \draw[black] (.7, -.7) node[anchor= west]{$\phi (k_2)$} to (.1, -.1);
        \draw[black] (-.1,-.1) to (-.6,-.6) to[out=-135, in=-90] (-.9,0);
        \draw[-{To}, black] (-.1, .1) to (-.6, .6) to[out= 135, in= 90] (-.9,0)
            node[anchor = west]{$k_3$};
        \draw (0,-.8) node[anchor = north]{$V(3312)$} to (0,-.8);
    \end{tikzpicture}
    \end{minipage}%
    \begin{minipage}{.249\textwidth}
         \centering
    \begin{tikzpicture}[scale = 1.2]
        \draw[black] (-.1,-.1) rectangle (.1,.1);
        \draw[black] (-.7, -.7) node[anchor= east]{$\phi (k_2)$} to (-.1, -.1);
        \draw[black] ( .7, -.7) node[anchor= west]{$\phib(k_1)$} to ( .1, -.1);
        \draw[-{To}, black] (-.1,  .1) to (-.6,  .6) to[out=135, in=180] (0, .9)
            node[anchor = north]{$k_3$};
        \draw[black] (.1,.1) to (.6,.6) to[out=45, in=0] (0, .9);
        \draw (0,-.8) node[anchor = north]{$V(3213)$} to (0,-.8);
    \end{tikzpicture}
    \end{minipage}%
    \begin{minipage}{.249\textwidth}
         \centering
    \begin{tikzpicture}[scale = 1.2]
        \draw[black] (-.1,-.1) rectangle (.1,.1);
        \draw[black] (-.7,.7) node[anchor= east]{$\phib(k_1)$} to (-.1,.1);
        \draw[black] ( .7,.7) node[anchor= west]{$\phi (k_2)$} to ( .1,.1);
        \draw[-{To}, black] (-.1, -.1) to (-.6, -.6) to[out=-135, in=180] (0, -.9)
            node[anchor = south]{$k_3$};
        \draw[black] (.1,-.1) to (.6,-.6) to[out=-45, in=0] (0, -.9);
        \draw (0,-.9) node[anchor = north]{$V(1332)$} to (0,-.9);
    \end{tikzpicture}
    \end{minipage}
    
    \caption{The four Feynman diagrams associated to the vertex functions of equation \eqref{cestlesdiagram}.}
    \label{fig:planar}
\end{figure}
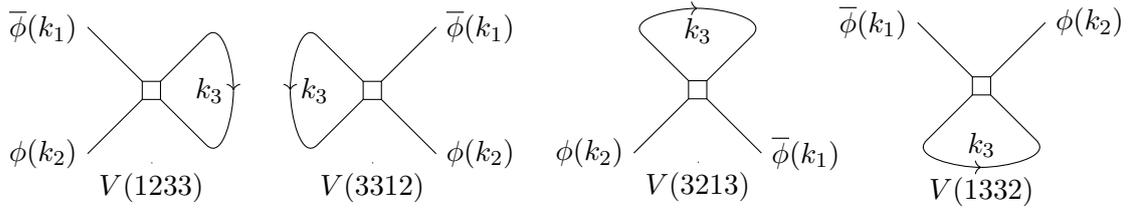

\paragraph{}
Now, one combines
\begin{equation}
    S_{\mathrm{int}}\left( \frac{\delta}{\delta J },\frac{\delta}{\delta\Jb} \right)
    = (2\pi)^4 \int \Big(\prod_{j=1}^4dk_j \Big)\ \frac{\delta}{\delta J(k_1)} \frac{\delta}{\delta\Jb(k_2)} \frac{\delta}{\delta J(k_3)} \frac{\delta}{\delta\Jb(k_4)}\ V(k_1, k_2, k_3, k_4),
    \label{csint-funct}
\end{equation}
with \eqref{generateur} and \eqref{Wzero}. After some algebra and making use of the Legendre transform $J(p) = (p^2 + m^2) \phi(p)$, $\Jb(p) = (p^2+m^2) \phib(p)$ to obtain the 1-loop contributions to the effective action whose general definition is
\begin{equation}
    \Gamma(\phi, \phib)
    = \int d^4k \Big(\Jb(k) \phi(k) + J(k) \phib(k) \Big) - W(J, \Jb),
\end{equation}
with
\begin{align}
    \phib(k) = \frac{\delta W(J, \Jb)}{\delta J  (k)}, &&
    \phi (k) = \frac{\delta W(J, \Jb)}{\delta \Jb(k)},
\end{align}
we are lead to the quadratic part of the one-loop effective action $\Gamma^{(2)}(\phi,\phib)$ given by
\begin{equation}
    \Gamma^{(2)}(\phi, \phib)
    = \int d^4k_1\ d^4k_2\ \phib(k_1) \phi(k_2) \Gamma^{(2)}(k_1, k_2),
\end{equation}
where, upon setting $V(1234):=V(k_1,k_2,k_3,k_4)$, one has
\begin{equation}
    \Gamma^{(2)}(k_1, k_2)
    = \int d^4k_3\ (k_3^2 + m^2)^{-1} \big(V(3312) + V(1233) + V(1332) + V(3213) \big).
    \label{cestlesdiagram}
\end{equation}
These 4 contributions, with external momenta $k_1$ and $k_2$, are pictured in Figure \ref{fig:planar}. These can be easily computed by simply dealing with the related delta functions. 

Note that using the symmetries \eqref{eq:vertex_symm}, one has $V(1233) = V(3312)$ and $V(3213) = V(1332)$ so that the four contribution of Figure \ref{fig:planar} crumbles down to two contributions.

\paragraph{}
For instance, pick the first contribution $V(3312)$ in \eqref{cestlesdiagram}, denoted hereafter by $\Gamma_1^{(2)}(k_1, k_2)$. From \eqref{zebigvertex}, one infers that the relevant delta's are
\begin{equation}
    \delta(k^0_1-k^0_2)\ \delta(k^3_1-k^3_2)\
    \delta^2\Big( R(\rho k^0_2) (\vec{k}_1 - \vec{k}_2) \Big)
\end{equation}
which, upon using the identity $\delta^2(R\vec{p})=|\det  R|^{-1}\delta^2{(\vec{p}})$, collapses to $\delta^4(k_1-k_2)$. Hence one obtains 
\begin{equation}
    \Gamma_1^{(2)}(k_1,k_2)
    \sim \int d^4k_3\ \frac{1}{k_3^2 + m^2}.
    \label{decadix}
\end{equation}
One would proceed in a similar way for the three other contribution, leading to the same result.

\paragraph{}
Let us discuss this result. In view of \eqref{decadix} and the above result, the one-loop 2-point function exhibits a UV quadratic divergence for the 4-dimensional theory, as its commutative counterpart to which it is similar. The corresponding contributions are related to planar diagrams so that they do not depend on the external momenta $k_1$ and $k_2$. Note that planar contributions to the 2-point function arising in the (real-)scalar field theory on $\rho$-Minkowski studied in \cite{rho3} also diverge as the commutative $\phi^4$ theory.

Besides, no IR singularities appear in the 2-point function which could generate UV/IR mixing. Only planar diagrams contribute to the 2-point function. Note that this can be expected within a theory involving a complex scalar field with orientable interaction.

A similar conclusion holds for the 3-dimensional case, with however the UV divergence being linear instead of the quadratic divergence in 4 dimensions.

\begin{figure}
    \begin{minipage}{.24\textwidth}
         \centering
    \begin{tikzpicture}[scale = 1.2]
        \draw[black] (-.1,-.1) rectangle (.1,.1);
        \draw[black] (-.7, -.7) node[anchor= east]{$\phi (k_4)$} to (-.1, -.1);
        \draw[black] (-.7,  .7) node[anchor= east]{$\phib(k_3)$} to (-.1,  .1);
        \draw[-{To}, black] (.1, .1) to (.6, .6) to[out= 45, in= 90] (.9,0)
            node[anchor = east]{$k_1$};
        \draw[black] (.1,-.1) to (.6,-.6) to[out=-45, in=-90] (.9,0);
        \draw (0,-.8) node[anchor = north]{$V^{\mathrm{no}}(3411)$} to (0,-.8);
    \end{tikzpicture}
    \end{minipage}%
    \begin{minipage}{.24\textwidth}
         \centering
    \begin{tikzpicture}[scale = 1.2]
        \draw[black] (-.1,-.1) rectangle (.1,.1);
        \draw[black] (.7,  .7) node[anchor= west]{$\phib(k_3)$} to (.1,  .1);
        \draw[black] (.7, -.7) node[anchor= west]{$\phi (k_4)$} to (.1, -.1);
        \draw[black] (-.1,-.1) to (-.6,-.6) to[out=-135, in=-90] (-.9,0);
        \draw[-{To}, black] (-.1, .1) to (-.6, .6) to[out= 135, in= 90] (-.9,0)
            node[anchor = west]{$k_1$};;
        \draw (0,-.8) node[anchor = north]{$V^{\mathrm{no}}(1134)$} to (0,-.8);
    \end{tikzpicture}
    \end{minipage}%
    \begin{minipage}{.26\textwidth}
         \centering
    \begin{tikzpicture}[scale = 1.2]
        \draw[black] (-.1,-.1) rectangle (.1,.1);
        \draw[black] (-.7, -.7) node[anchor= east]{$\phi (k_4)$} to (-.1, -.1);
        \draw[black] ( .7,  .7) node[anchor= west]{$\phib(k_3)$} to ( .1,  .1);
        \draw[black,
            decoration={markings, mark=at position 0.75 with {\arrow{To}}},
            postaction={decorate}
        ] (-.1,  .1) to (-.6,  .6) to[out=135, in=135] (.45, .55);
        \node at (0,1) {$k_1$};
        \draw[black] ( .1, -.1) to ( .6, -.6) to[out=-45, in=-45] (.55, .45);
        \draw (0,-.8) node[anchor = north]{$V^{\mathrm{no}}(1431)$} to (0,-.8);
    \end{tikzpicture}
    \end{minipage}%
    \begin{minipage}{.26\textwidth}
         \centering
    \begin{tikzpicture}[scale = 1.2]
        \draw[black] (-.1,-.1) rectangle (.1,.1);
        \draw[black] (-.7,  .7) node[anchor= east]{$\phib(k_3)$} to (-.1,  .1);
        \draw[black] ( .7, -.7) node[anchor= west]{$\phi (k_4)$} to ( .1, -.1);
        \draw[black,
            decoration={markings, mark=at position 0.75 with {\arrow{To}}},
            postaction={decorate}
        ] ( .1,  .1) to ( .6,  .6) to[out=  45, in=  45] (.55, -.45);
        \node at (1,0) {$k_1$};
        \draw[black] (-.1, -.1) to (-.6, -.6) to[out=-135, in=-135] (.45, -.55);
        \draw (0,-.8) node[anchor = north]{$V^{\mathrm{no}}(3114)$} to (0,-.8);
    \end{tikzpicture}
    \end{minipage}
    
    \caption{The four Feynman diagram associated to the vertex functions of equation \eqref{cestlesdiagram-bis}.}
    \label{fig:no_vertex}
\end{figure}
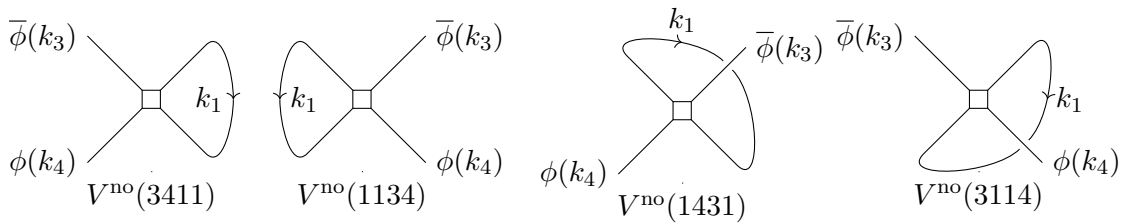

\paragraph{}
It is instructive to study the 1-loop behaviour of the 2-point function that would arise for a noncommutative complex scalar theory with a non-orientable interaction replacing the quartic term in \eqref{action1}. For that purpose, we will consider an interaction term of the form 
\begin{equation}
    S^{\mathrm{no}}_{\mathrm{int}}
    = g \langle \phi \star_\rho \phi, \phi \star_\rho \phi \rangle
    = g \int d^4x\ (\phi^\dag \star_\rho \phi^\dag \star_\rho \phi \star_\rho \phi)(x),
    \label{nonorient-1}
\end{equation}
while the quadratic part of \eqref{action1} remains unchanged. Then, the analysis performed above can be thoroughly reproduced, the only change being the replacement of the vertex function \eqref{zebigvertex} by
\begin{align}
\begin{aligned}
    {V}^{\mathrm{no}}(k_1, k_2, k_3, k_4)
    =&\; g\ \delta(k^0_1 - k^0_2 + k^0_3 - k^0_4) \delta(k^3_1 - k^3_2 + k^3_3 - k^3_4)\\
    &\times \delta^2 \Big(\vec{k}_1 - \vec{k}_2 + R(\rho k^0_1) \vec{k}_3 - R(\rho k^0_2) \vec{k}_4 \Big).
\end{aligned}
    \label{zebigvertex-bis}
\end{align}
The resulting 2-point function at one-loop takes the form
\begin{equation}
    \Gamma^{(2)}(\phi, \phib)
    = \int d^4k_3\ d^4k_4\ \phib(k_3) \phi(k_4)\ \Gamma^{(2)\mathrm{no}}(k_3, k_4),
\end{equation}
with
\begin{equation}
\begin{aligned}
    &\Gamma^{(2)\mathrm{no}} (k_3, k_4) \\
    &= \int d^4k_1\ (k_1^2 + m^2)^{-1} \Big(V^{\mathrm{no}}(1134)  + V^{\mathrm{no}}(3411) + V^{\mathrm{no}}(3114) + V^{\mathrm{no}}(1431) \Big).
\end{aligned}
    \label{cestlesdiagram-bis}
\end{equation}
This expression is actually similar to \eqref{cestlesdiagram}, upon replacing $V$ by $V^{\mathrm{no}}$. Its vertices are represented in Figure \ref{fig:no_vertex}. By combining the delta's appearing in the vertex functions, it can be easily seen that the two first contributions in \eqref{cestlesdiagram-bis} do not depend on the external momenta $k_3$ and $k_4$ and thus corresponds to planar diagrams, proportional to $\int d^4p\ (p^2+m^2)^{-1}$ and thus are UV quadratically diverging (which become linearly diverging for 3-dimensional case).

Note that the symmetry \eqref{eq:vertex_symm} does not hold for the vertex $V^{\mathrm{no}}$ \eqref{zebigvertex-bis}, so that \textit{a priori} none of these diagrams could be equalized by symmetry arguments.

\paragraph{}
Unlike the planar contributions, the last two contributions depend on the external momenta. They correspond to non-planar diagrams whose contributions may become singular at zero external momenta, thus generating UV/IR mixing. Note that these latter diagrams are somewhat comparable to the so-called type IV diagrams arising in non-orientable scalar field theories on $\kappa$-Minkowski space \cite{PW2018}.

\paragraph{}
To illustrate the appearance of IR singularities, consider for instance the third contribution in \eqref{cestlesdiagram-bis}.  We assume $\rho\ne0$ in the following. From \eqref{zebigvertex-bis}, one can verify that the delta's in the vertex function are 
\begin{equation}
    \delta^2 \Big(\vec{k}_3 - \vec{k}_1 + R(\rho k^0_3) \vec{k}_1 - R(\rho k^0_1) \vec{k}_4 \Big)\
    \delta(k^0_3 - k^0_4)\ \delta(k^3_3 - k^3_4). 
\end{equation}
Then, the integration over $\vec{k}_1$ produces 
\begin{equation}
    \Gamma_3^{(2)\mathrm{no}}(k_3,k_4)
    \sim \int dk^0_1\ \frac{1}{(k^0_1)^2 + K_1^2(\vec{k}_3, \vec{k}_4, k^0_1)} \ \big| \det(\mathcal{R}(k^0_3) \big|^{-1}
    \label{singlu-mix}
\end{equation}
where $K_1^2(\vec{k}_3,\vec{k}_4,k^0_1)$ is a function whose expression can be read off from the corresponding delta and we have set
\begin{equation}
   \mathcal{R}(k^0_3)
   = R(\rho k^0_3) - \bbone_2,
\end{equation}
which obviously vanishes for $k^0_3=0$ thus generating an IR singularity in \eqref{singlu-mix} which cannot be compensated by the remaining integral over $k^0_1$. This therefore signals that complex-scalar field theories on $\rho$-Minkowski space with non-orientable interaction term generally have UV/IR mixing. Note that the real-scalar field theory on $\rho$-Minkowski studied in \cite{rho3} exhibits necessarily non-planar contributions
to the 2-point function which are also IR singular so that both results agree.

\paragraph{}
As far as 2-point functions are concerned, one concludes that the UV behaviour of the 2-point functions in both scalar theories qualitatively agree.
Besides, we note that the status of the complex-scalar field theories on $\rho$-Minkowski space is globally similar to the one for their homologs on $\kappa$-Minkowski. Recall that among these latter, one family with orientable interaction was shown to have a vanishing beta function at the one-loop order, with corresponding one-loop corrections being UV finite \cite{PW2019}. This was due to the particular nature of the interaction vertex combined with the behaviour of the propagator exhibiting a rather strong UV decay.

\paragraph{}
For the sake of comparison, one interesting issue to investigate is the behaviour of the one-loop 4-point function within the {\it{orientable}} complex-scalar field theories on $\rho$-Minkowski space to which we turn now on.

\begin{figure}
    \begin{minipage}{.495\textwidth}
         \centering
    \begin{tikzpicture}[scale = 1.2]
        \draw[black] (-.8,-.1) rectangle (-.6,.1);
        \draw[black] ( .6,-.1) rectangle ( .8,.1);
        \draw[black] (-1.2, -.7) node[anchor= east]{$\phi (k_2)$} to (-.8, -.1);
        \draw[black] (-1.2,  .7) node[anchor= east]{$\phib(k_1)$} to (-.8,  .1);
        \draw[black] ( 1.2, -.7) node[anchor= west]{$\phib(k_3)$} to ( .8, -.1);
        \draw[black] ( 1.2,  .7) node[anchor= west]{$\phi (k_4)$} to ( .8,  .1);
        \draw[black,
            decoration={markings, mark=at position .5 with {\arrow{To}}},
            postaction={decorate}
            ] (-.6,  .1) to[out=45, in=135] (.6, .1);
        \node at (0,.6) {$k_5$};
        \draw[black,
            decoration={markings, mark=at position .5 with {\arrow{To}}},
            postaction={decorate}
            ] ( .6,  -.1) to[out=-135, in=-45] (-.6, -.1);
        \node at (0,-.6) {$k_6$};
        \draw (0,-1) node[anchor = north]{$V(1256)V(3465) = V(1256)V(5643)$} to (0,-1);
    \end{tikzpicture}
    \end{minipage}%
    \begin{minipage}{.495\textwidth}
         \centering
    \begin{tikzpicture}[scale = 1.2]
        \draw[black] (-.8,-.1) rectangle (-.6,.1);
        \draw[black] ( .6,-.1) rectangle ( .8,.1);
        \draw[black] (-1.2, -.7) node[anchor= east]{$\phi (k_4)$} to (-.8, -.1);
        \draw[black] (  .2,  .7) node[anchor= south west]{$\phib(k_3)$} to ( .6,  .1);
        \draw[black] ( -.2,  .7) node[anchor= south east]{$\phi (k_2)$} to (-.6,  .1);
        \draw[black] ( 1.2,  .7) node[anchor= west]{$\phib(k_1)$} to ( .8,  .1);
        \draw[black] (-.8,.1) to (-1.1, .6) to[out=135, in=135] (-1.1, -.5);
        \draw[black,
            decoration={markings, mark=at position .5 with {\arrow{To}}},
            postaction={decorate}
            ] (-1.05, -.55) to[out=-45, in=-45] (1.1, -.4) to (.8, -.1);
        \node at (.4,-.7) {$k_5$};
        \draw[black,
            decoration={markings, mark=at position .5 with {\arrow{To}}},
            postaction={decorate}
            ] ( .6,  -.1) to[out=-135, in=-45] (-.6, -.1);
        \node at (0,0) {$k_6$};
        \draw (0,-1) node[anchor = north]{$V(5462)V(3615)$} to (0,-1);
    \end{tikzpicture}
    \end{minipage}%
    
    \caption{The two Feynman diagrams associated to the vertex \eqref{4-point} and \eqref{4-point-planar}.}
    \label{fig:4_pt}
\end{figure}
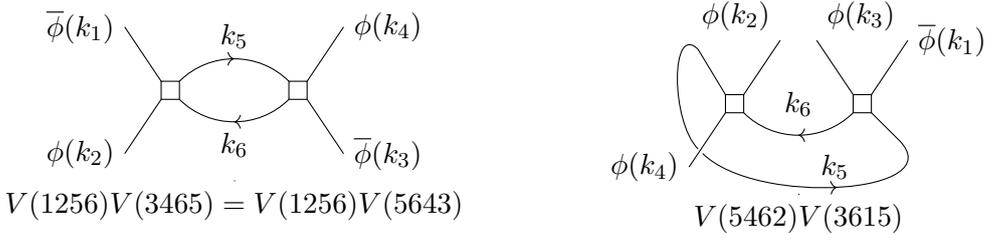

\paragraph{}
To address this problem, consider the following diagram contribution, among the twelve planar and non-planar contributions to the 4-point function
\begin{equation}
    \Gamma^{(4)}(k_1, k_2, k_3, k_4)
    \sim \int d^4k_5\ d^4k_6\ \frac{1}{(k^2_5 + m^2)(k^2_6 + m^2)} V(5462) V(3615),
    \label{4-point}
\end{equation}
in which $V$ is still given by \eqref{zebigvertex} and the external momenta are $k_1, k_2, k_3, k_4$.

The integration over $d^4k_6$ yields
\begin{align}
\begin{aligned}
    \Gamma^{(4)} (k_1, k_2, & k_3, k_4) \\
    \sim \int & d^4k_5\ \frac{1}{(k^2_5 + m^2)} \frac{1}{(k^0_5 - k^0_2 - k^0_4)^2 + \big(R_{5-2}(\vec{k}_5 - \vec{k}_4) + \vec{k}_2 \big)^2 + m^2} \\
    &\times \delta(k^0_1 - k^0_2 + k^0_3 - k^0_4)\ \delta(k^3_1 - k^3_2 + k^3_3 - k^3_4) \\
    &\times \delta^2 \Big(\vec{k}_3 - \vec{k}_2 + R_{5-2}(\vec{k}_5 - \vec{k}_4) + R_{5-3}(\vec{k}_1 - \vec{k}_5) \Big),
\end{aligned}
    \label{stuff1}
\end{align}
where we set 
\begin{align}
    R_{j-l} := R \big(\rho (k^0_j - k^0_l) \big), &&
    R_j := R(\rho(k^0_j)).   
\end{align}
Observe that the first two delta's express the ("undeformed") conservation of the energy and the third component of the external momenta while the last one involves $\vec{k}_5$ whose value will be uniquely fixed as a function of the external momenta (and $k^0_5$) upon integrating over $\vec{k}_5$, except possibly for exceptional external momenta, as shown below.

\paragraph{}
Now, the integration over $d\vec{k}_5$ leads to
\begin{align}
\begin{aligned}
    \Gamma^{(4)}(k_1, k_2, k_3, k_4)
    \sim \int & dk^0_5\ \frac{1}{(k^0_5)^2 + (\vec{k}_5^\star)^2 + m^2} 
    \times\frac{1}{\big|1 - \cos \big(\rho(k^0_3 - k^0_2) \big) \big|} \\
    &\times \frac{1}{(k^0_5 - k^0_2 - k^0_4)^2 + \big(R_{5-2}(\vec{k}_5^\star - \vec{k}_4) + \vec{k}_2 \big)^2 + m^2}
\end{aligned}
    \label{decadix1}
\end{align}
in which we have omitted the two delta's related to the "undeformed" conservation laws and $\vec{k}_5^\star$ is solution of 
\begin{equation}
    (R_{3-2} - \mathbb{I}_2) \vec{k}_5^\star
    = R_{3-2} \vec{k}_4 + R_{3-5}(\vec{k}_2 - \vec{k}_3) - \vec{k}_1 ,
\end{equation}
whenever $k^0_3-k^0_2\ne0$. Assume this condition holds. Then, at large $k^0_5$, one infers that $\vec{k}_5^\star$ behaves as a function of the only external momenta. Hence, it can be seen that the integrand behave as $\sim \left( \frac{1}{k^0_5} \right)^4$ so that the remaining integral in \eqref{decadix1} is UV finite.

Furthermore, in view of the last factor in \eqref{decadix1}, one concludes that the diagram is singular when $k^0_3-k^0_2=0$ and in particular when both momenta are vanishing. Therefore, dangerous IR singularities generating UV/IR mixing occur at one-loop in this contribution to the 4-point function.

\paragraph{}
Consider now the following planar diagram contribution 
\begin{equation}
    \Gamma_P^{(4)}(k_1, k_2, k_3, k_4)
    \sim \int d^4k_5\ d^4k_6\ \frac{1}{(k^2_5 + m^2) (k^2_6 + m^2)} V(1256) V(3465).
    \label{4-point-planar}
\end{equation}
By performing a computation similar to the one given just above, one obtains
\begin{equation}
\begin{aligned}
    \Gamma_P^{(4)} (k_1, k_2, & k_3, k_4) \\
    \sim \int & d^4k_5\ \frac{1}{k_5^2 + m^2}\ \frac{1}{(k_1^0 - k_2^0 + k^0_5)^2 + \Big(\vec{k}_5 + R_{2-5}(\vec{k}_1 - \vec{k}_2) \Big)^2 + m^2} \\
    & \times \delta(k^0_1 - k^0_2 + k^0_3 - k^0_4)\ \delta(k^3_1 - k^3_2 + k^3_3 - k^3_4) \\
    & \times \delta^2 \big( R_3(\vec{k}_3 - \vec{k}_4) + R_2(\vec{k}_1 - \vec{k}_2) \big).
\end{aligned}
    \label{planar-final}
\end{equation}
Compared to \eqref{stuff1} where one delta function still involves the internal momentum which thus will be (partly) fixed when integrated over, leading to an UV finite expression, eqn. \eqref{planar-final} involves one remaining integration over $d^4k_5$ while the delta's depend only on the external momenta.

By a simple inspection, one then easily realizes that the integral is logarithmically diverging in 4 dimensions. This is nothing but the UV behaviour of the commutative $\phi^4$ theory. Note that this analysis can be straightforwardly extended to the 3-dimensional case, leading to a UV finite contribution.

\paragraph{}
Although the full computation of the 1-loop corrections to the 4-point function is beyond the scope of this paper, we may already indicates what would likely come out from such a computation.

The main point is that one can expect that singularities as the one exhibited in \eqref{decadix1} will likely occur in the various diagrams. Furthermore, there is no apparent reason why they might balance each other so that UV/IR mixing is expected to plague the scalar theory with orientable interaction, originated by (IR) singularities in the one-loop 4-point function. We regard the singularity in \eqref{decadix1} as being already a sufficient evidence for the occurrence of the mixing.

The UV behaviour of the model is similar to the one of the ordinary $\phi^4$ theory. Such a result is not surprising regarding the UV decay property of the propagator of \eqref{action1} which is the ordinary one. Note that other kinetic operators built from various noncommutative differential calculus and having a faster UV decay may obviously give rise to UV finitude in 4 dimensions.

\section{Conclusion}
\label{sec:section4}
\paragraph{}
We have studied one-loop perturbative properties of complex-valued scalar field theories on the $\rho$-Minkowski space. The corresponding star-product is different from the one used in \cite{marija1}, \cite{rho3} based on a Drinfeld twist or a Jordan-Schwinger map \cite{infinimany}. It is obtained by adapting the construction used in the case of the $\kappa$-Minkowski space \cite{DS, PW2018} inherited from the old works \cite{vonNeum, Weyl}: the defining items of the convolution algebra of the Lie group linked to the coordinate algebra of $\rho$-Minkowski are transferred by the Weyl quantization to the star-product, the involution and the natural integration measure and trace which is not twisted. This characterizes the associative algebra modeling $\rho$-Minkowski.

\paragraph{}
One-loop properties for the 2-point and 4-point functions are examined. Four dimensional scalar theory with quartic orientable interaction has one-loop UV quadratically diverging 2-point function while no IR singularities generating UV/IR mixing appears. Theory with quartic non-orientable interaction has a 2-point function plagued with IR singularity thus generating UV/IR mixing.

The 4-point function for the theory with orientable interaction receives UV logarithmically diverging contributions. Furthermore, it involves IR singularities which thus signals the appearance of UV/IR mixing.

\paragraph{}
While \cite{rho3} and the present work provide a better insight on the landscape of field theories on $\rho$-Minkowski space, two immediate issues should now be examined. One concerns the effects of the changes of kinetic operator (propagator) on the UV as well as IR behaviour. The other one concerns the elaboration of a gauge theory model on $\rho$-Minkowski. It turns out that both issues are obviously related to the types of noncommutative differential calculus which come into play. On a more algebraic viewpoint, it would be interesting to identify a twist giving rise to the star-product used in the present study. We will return to these questions in forthcoming publications.

\section*{Acknowledgments}
We thank P.\ Vitale for discussions on $\rho$-Minkowski space and related noncommutative field theories. J.-C.\ W is grateful to A.\ Wallet for numerous discussions on various aspects of harmonic analysis of locally compact groups.

We thanks the Action CA18108 QG-MM ``Quantum Gravity phenomenology in the multi-messenger approach'' and the Action 21109 CaLISTA ``Cartan geometry, Lie, Integrable Systems, quantum group Theories for Applications'', from the European Cooperation in Science and Technology.

\end{document}